\documentclass[twocolumn,secnumarabic,amssymb, nobibnotes, aps,prl]{revtex4-1}
\pdfoutput=1
\usepackage{graphicx}

\newcommand{\sev}{7$\times$7 } 

\begin{document}

\title{A phantom force induced by the tunneling current, characterized on Si(111)}
\date{\today}

\author{A.J.~Weymouth}
\author{T.~Wutscher}
\author{J.~Welker}
\author{T.~Hofmann}
\author{F.J.~Giessibl}
\affiliation{Institute of Experimental and Applied Physics, University of Regensburg, D-93053 Regensburg, Germany.}

\begin{abstract}
Simultaneous measurements of tunneling currents and atomic forces on surfaces and 
adsorbates 
provide new insights into the electronic and structural properties of matter on the atomic scale.
We report on experimental observations and calculations of a strong impact the tunneling current can have on the measured force, 
which arises
when the resistivity of the sample cannot be neglected.
We present a study on Si(111)--\sev with various doping levels, but this effect is expected to occur on other low-conductance samples like adsorbed molecules, and is likely to strongly affect Kelvin probe measurements on the atomic scale.

\end{abstract}

\maketitle
%What needs to be stressed is simultaneous force and current measurements
Scanning tunneling microscopy (STM) sparked enthusiasm in scanning probe microscopy with images of the adatoms of Si(111)--\sev~\cite{Binnig1983}.
The atomic force microscope (AFM) removed the requirement for a conducting substrate~\cite{Binnig1986} but brought more than just the possibility to measure on insulators~\cite{Giessibl1992}: 
Subatomic~\cite{Giessibl2006} and submolecular~\cite{Gross2009} imaging have also been demonstrated.
These successes have brought strong interest in combined STM and AFM (e.g. \cite{Ternes2011}), however the independence of force and current measurements remains an open issue~\cite{Molitor1999, Arai2000, Arai2000a, Guggisberg2001, Sugimoto2010}.

Frequency-modulation AFM (FM-AFM) is a technique in which the interaction between tip and sample is measured by the frequency shift, $\Delta f$, of an oscillating tip from its eigenfrequency, $f_0$~\cite{Albrecht1990}.
$\Delta f$ can be formulated as a measure of the force gradient, $k_{ts}=-\frac{\mathrm{d}F}{\mathrm{d}z}$, where $z$ is the distance from the surface.
$\Delta f$ is also a function of the spring constant of the oscillator, $k$, the amplitude of oscillation, $A$, and $z$, and can be 
approximated at small amplitudes by $\Delta f \approx (f_0 / 2\,k)\,k_{ts}$~\cite{Giessibl2001}.
In short, a decrease in $\Delta f$ indicates that the force between the tip and sample is becoming more attractive.

%%%%%%%%%%%%%%%%%%%%%%%%%%%%%%%%%%%%%%%%%%%%%%%
\begin{figure}%[h]
\begin{center}
\includegraphics[width=\columnwidth]{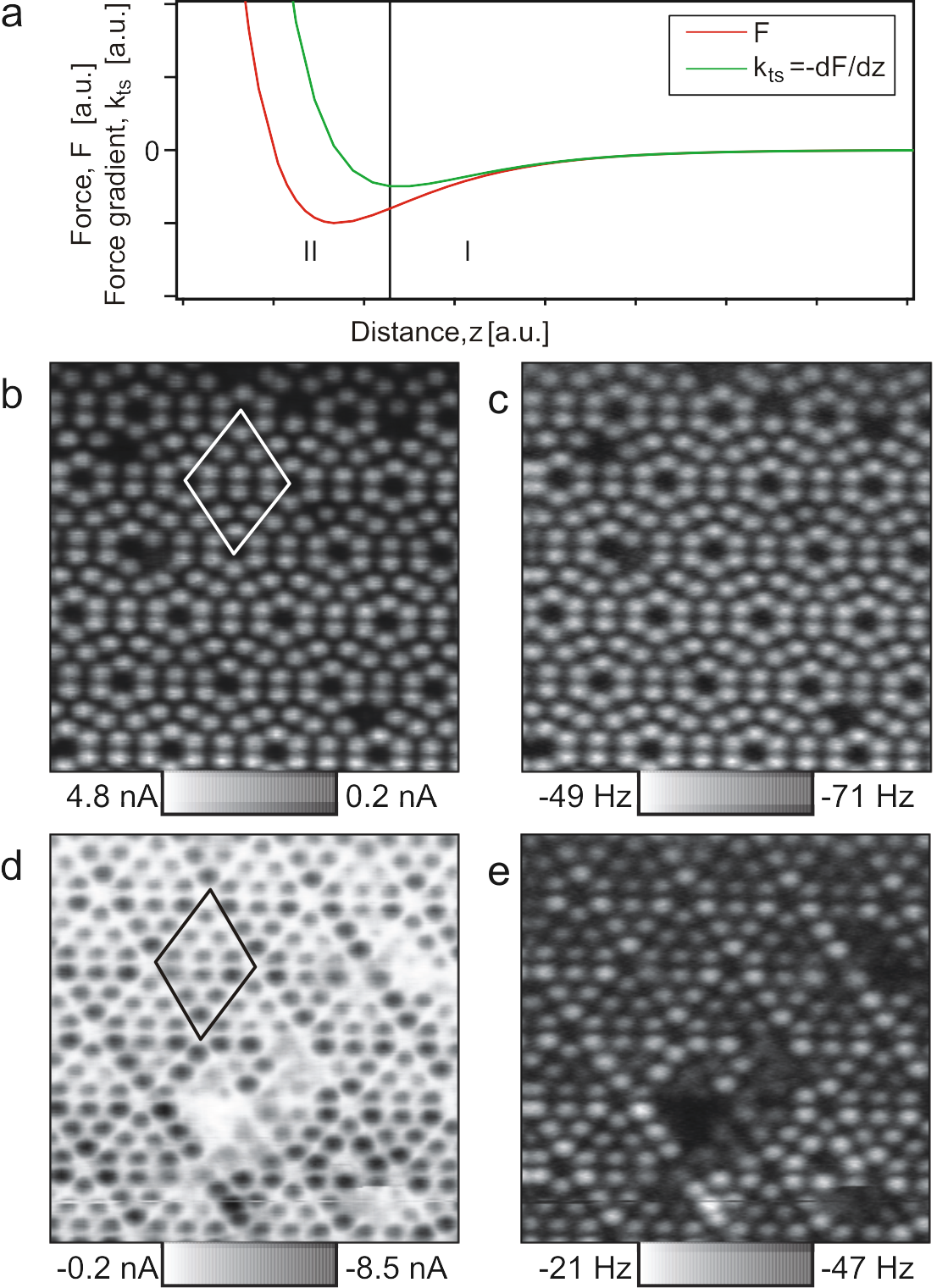}

\caption{
a) From a Morse potential, the force and force gradient can be calculated.  
At $V_{\mathrm{tip}}=-1.5$\,V, simultaneous b) current and c) force data are collected without $I$ or $\Delta f$ feedback.  Similar data are collected at $V_{\mathrm{tip}}=+1.5$\,V; the d) current data appears to be inverted because in opposite bias voltages, current flow is reversed, however e) $\Delta f$ again increases above adatoms.  Data were collected at $A=400$\,pm, $f_0=25\,908\,$Hz.  In STM data, a unit cell of \sev is highlighted; images are 10\,nm$\times$10\,nm. 
}
\label{si111}
\end{center}
\end{figure}
%%%%%%%%%%%%%%%%%%%%%%%%%%%%%%%%%%%%%%%%%%%%%%

In contrast to the tunneling current, $I$, measured with STM, $\Delta f$ is not monotonic as a function of $z$.
The local tip-sample interaction is usually well-represented by a Morse potential from which the force and the force gradient can  be derived, as shown in Fig.~\ref{si111}(a).
In region I, $k_{ts}$ decreases as $z$ decreases.
Long-range forces (e.g. van der Waals) cause attractive interaction between tip and sample.
It is in this region STM is usually conducted on Si(111)--7$\times$7, at setpoints under 10\,nA at 1\,V, corresponding to tip-sample distances greater than 5\,\AA~\cite{Jelinek2008}. 
In region II, $k_{ts}$ increases as $z$ decreases.  
The waveform overlap between tip and sample causes measurable energy increase due to Pauli repulsion, which states electrons may not occupy the same quantum state~\cite{Moll2010}.

In this Letter, we report upon the effect of bias voltage on FM-AFM of Si(111)--7$\times$7.
At tip-sample distances corresponding to normal STM setpoints, one expects a decrease in frequency shift as the tip moves laterally without feedback over an adatom, due to the increase in attractive force~\cite{Perez1997}.
However, with the application of a moderate bias voltage ($>$1.0\,V), one is able to observe a frequency shift increase as the tip moves over an adatom.
Moreover, FM-AFM images taken with this applied bias voltage can show atomic contrast at tip-sample distances 300\,pm further from the surface than is required to image with no applied bias. 
We propose a model incorporating sample resistance where the observed frequency shift is caused by a decrease in the electrostatic attraction between tip and sample.

Experiments were performed with a qPlus sensor with $k=1800\,$N\,m$^{-1}$.
Data were collected in constant height mode with both a home-built microscope operating at room temperature in UHV and, where explicitly stated, at 4.2\,K with an Omicron LT-SPM.
Two types of Si(111) samples were used: a high-doped sample corresponding to a resistivity $\rho=0.010-0.012\,\Omega\,$cm at 300\,K and a low-doped sample with $\rho=6-9\,\Omega\,$cm at 300\,K.
Si(111)--\sev was prepared with several flash and anneal cycles.

%%%%%%%%%%%%%%%%%%%%%%%%%%%%%%%%%%%%%%%%%%%%%%%
\begin{figure}%[hb]
\begin{center}
%0.43  0.85
\includegraphics[width=0.85\columnwidth]{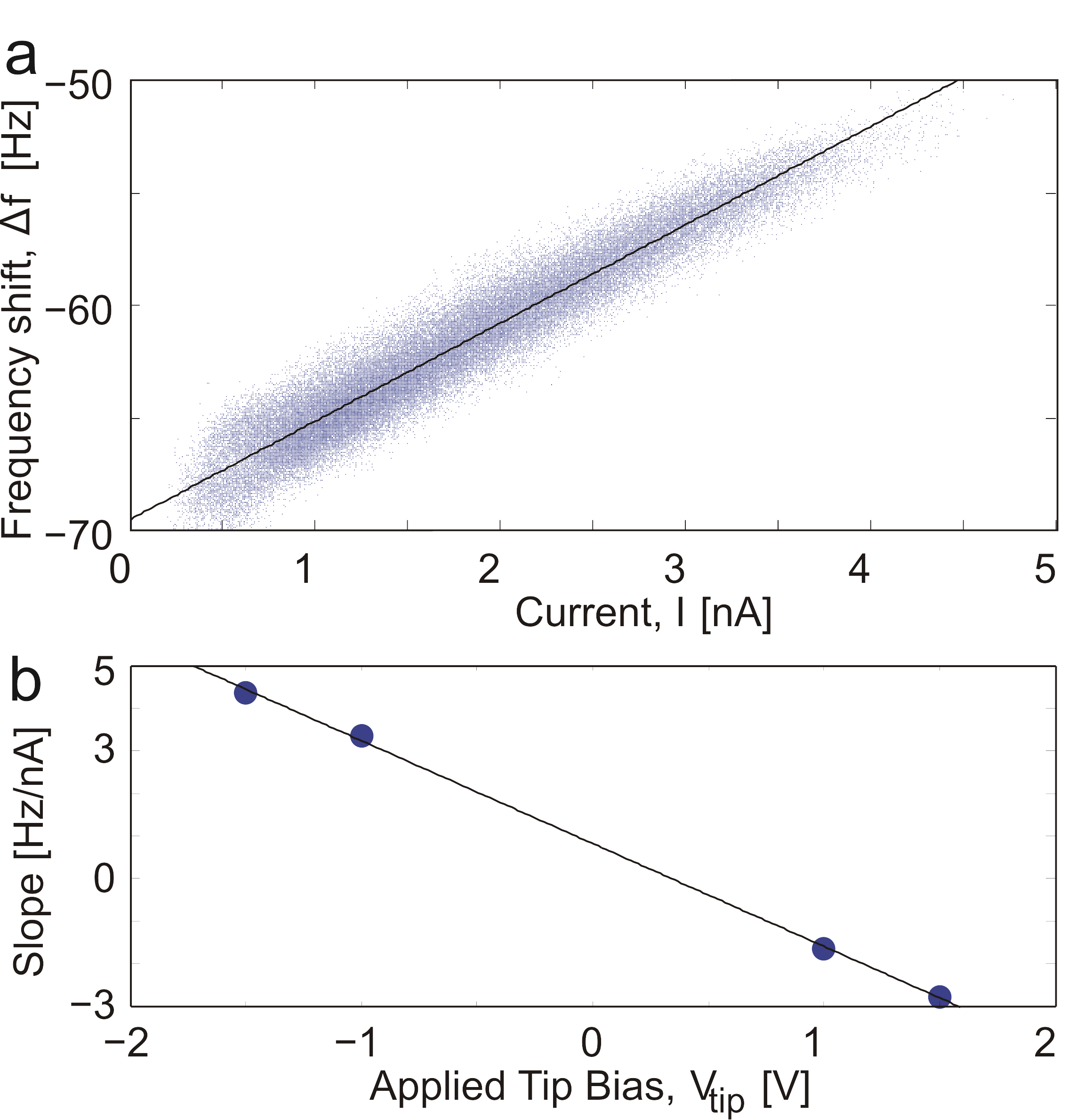}
\caption{
a) From Fig~\ref{si111}(b) and (c), a comparison between $I$ and $\Delta f$ can be made.  Their relationship is linear; the line shown has a slope of $4.36\,\mathrm{Hz}/ \mathrm{nA}$.  b) Following similar analyses at other biases ($V_{\mathrm{tip}}$=-1.5, -1.0, 1.0 and 1.5V) the slopes of the fits can be plotted as a function of the bias.  $I$ has the opposite sign as $V_{\mathrm{tip}}$.  A linear fit is shown as a guide to the eye.
}
\label{idf}
\end{center}
\end{figure}
%%%%%%%%%%%%%%%%%%%%%%%%%%%%%%%%%%%%%%%%%%%%%%%

\begin{figure}%[ht]
\begin{center}
\includegraphics[width=\columnwidth,clip]{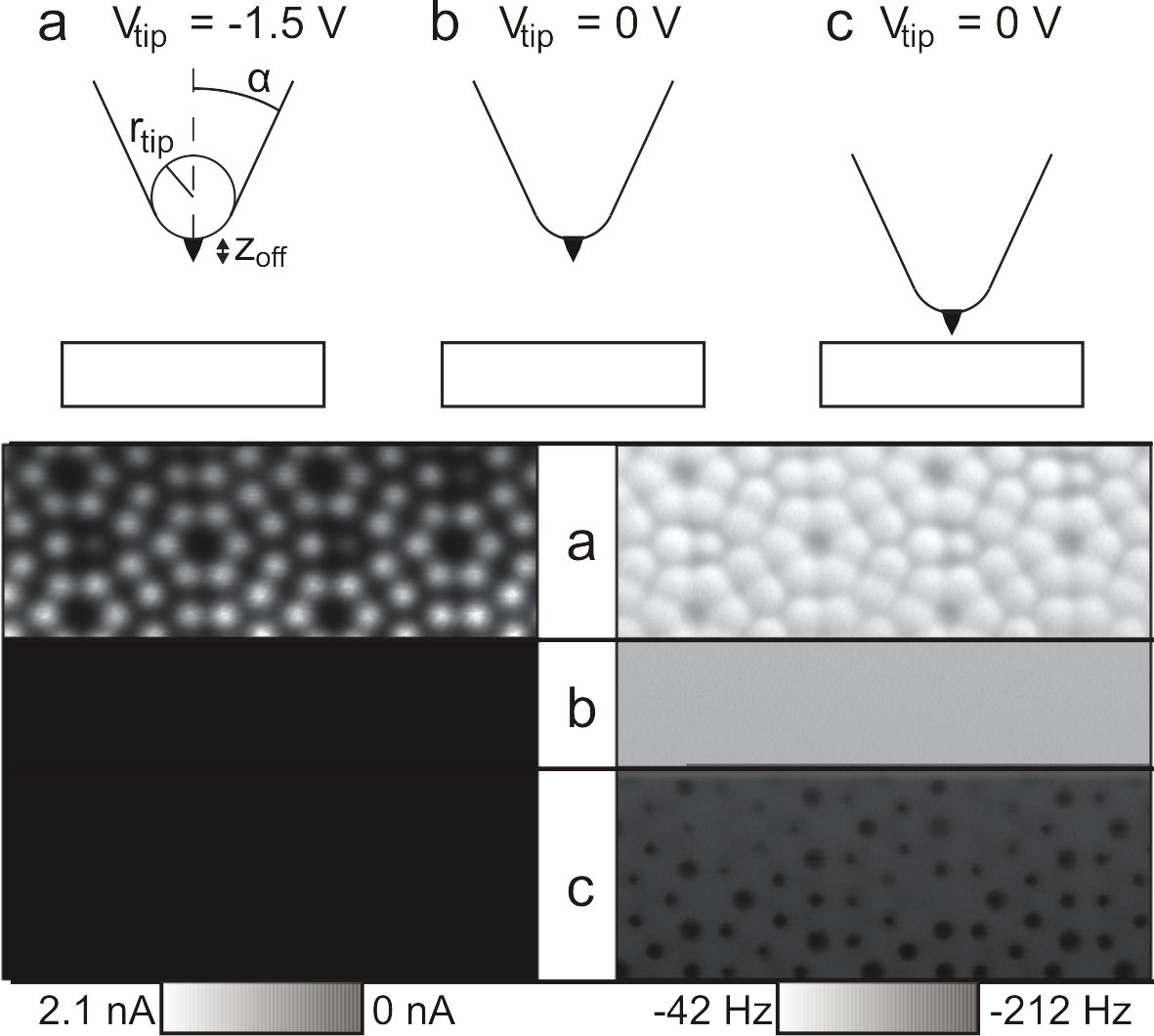}
\caption{
\textit{Top:} Schematic of the experiment.
\textit{Bottom:} Simultaneous $I$ and $\Delta f$ data acquired at 4.2\,K.  
a) $V_{\mathrm{tip}}=-1.5\,$V.  b) $V_{\mathrm{tip}}=0\,$V.  c) Tip is approached 340\,pm closer to the surface.  Data were collected at $A=100\,$pm, $f_0=16\,777\,$Hz and images are 8\,nm$\times$8\,nm.
}
\label{stm2}
\end{center}
\end{figure}

Figure \ref{si111}(b) to (e) show simultaneously acquired $I$ and $\Delta f$ data of the low-doped Si(111) sample.  
In (b) and (c), the tip bias was $V_{\mathrm{tip}}=-1.5$\,V.  
STM data show the clear structure of the \sev reconstruction, with all adatoms in the unit cell having approximately the same intensity.
The $\Delta f$ data show an increase in frequency shift above adatoms.
In (d) and (e), the tip bias was $V_{\mathrm{tip}}=+1.5$\,V.
The STM data show different features now in the outlined unit cell: over adatoms in the faulted half (the lower six adatoms), we record greater absolute current than those in the unfaulted half, as expected~\cite{HamersPRL1986}.
The $\Delta f$ data, however, also have stronger contrast above the faulted half unit cell. 
In a simple picture of AFM in which the total electron density is measured~\cite{Hembacher2004}, one would not expect this $\Delta f$ contrast to depend upon bias voltage. 
Also, while one might initially propose that this increase in $\Delta f$ over adatoms indicates that we are in region II of the Morse potential, we show later (with respect to the data in Fig.~\ref{stm2}) that this is not the case.

The relation between $\Delta f$ and $I$ channels can be further characterized.
For each pixel in image Fig.~\ref{si111}(b) to (e), force and current data have been acquired.  
Before a pixel-by-pixel comparision of $I$ and $\Delta f$ can be made, however, the low bandwidth of the PLL must be taken into account.  
At scan speeds even as low as 20\,nm\,s$^{-1}$, it can cause a noticable lateral displacement between $\Delta f$ and $I$ data.
Consider a scan line 10\,nm long with 256 pixels.
The bandwidth of our PLL is 120\,Hz.
Assuming the $I$ data to be instantaneous, the $\Delta f$ data will be offset by 
(256\,pixels/10\,nm) $\cdot$ 20\,nm\,s$^{-1}$ / 120\,Hz $\approx 4$ pixels.
Independently, a cross correlation of Fig.~\ref{si111}(b) and (c) show a 4 pixel offset.  

Each measurement of $I$ thus has an associated $\Delta f$ measurement, and these are plotted (for the $\Delta f$ and $I$ data shown in Fig.~\ref{si111} (b) and (c)) in Fig.~\ref{idf}(a). 
The fact that there is some correspondance does not come as a surprise, as regularly force and current images of the same surface appear similar.
What is surprising is that the data are quite linear and yield an increasing frequency shift over the entire current range.

The slope of the linear fit to the data in Fig.~\ref{idf}(a) is a measure of the $\Delta f$ response as a function of $I$.
This analysis was repeated with images at various biases ($V_{\mathrm{tip}}=-1.5\,$V, -1.0\,V, 1.0\,V and 1.5\,V).  
Fig. 3(b) shows the slopes of these fits with respect to applied biases.
The observation is that not only does this repulsive $\Delta f$ signal scale linearly with current, but that this $\Delta f$/$I$ slope itself scales with applied bias voltage.

%%%%%%%%%%%%%%%%%%%%%%%%%%%%%%%%%%%%%%%%%%%%%%%
%%%%%%%%%%%%%%%%%%%%%%%%%%%%%%%%%%%%%%%%%%%%%%%

We then performed measurements with the high-doped sample.  
At room temperature, this $\Delta f$ contrast was not observable in the current range of $|I|<5\,$nA.
At 4.2\,K, however, it was, as shown in Fig.~\ref{stm2}(a):
In cooling the system, the effective resistance increased and this $\Delta f$ contrast was again easily observed.

%This suggests that the $\Delta f$ increase is related to the effective resistance of the sample, which we expand upon later.
In order to investigate the tip-sample distance, the bias was reduced to zero partway through image acquisition.
The result is shown in Fig.~\ref{stm2}(b).
In this case, the contrast in $\Delta f$ disappears with the lack of an applied bias voltage.
Atomic contrast in $\Delta f$ should be observable without an applied bias voltage~\cite{Giessibl1995}.
In order to observe $\Delta f$ contrast with no applied bias voltage, we needed to advance the tip 340\,pm closer to the surface.
The remainder of the image, in Fig.~\ref{stm2}(c) on, clearly shows the expected contrast in $\Delta f$.

One might ask what effect removal of the bias voltage has upon the average tip-sample distance, knowing that there is an electrostatic force between tip and sample that scales as the square of the voltage difference~\cite{Hudlet1998}.
We have found, in agreement with previous studies~\cite{Guggisberg2000}, that a model of the tip that incorporates a spherical apex with a conic structure to account for long-range electrostatic interactions and a small nano-tip with negligable electrostatic contributions but with which imaging is performed, as shown in Fig.~\ref{stm2}(a), is quite accurate when describing long-range electrostatic forces.
Typical values to describe our tips are $r_{\mathrm{tip}}=5\,$nm, $\alpha=70^{\circ}$, and $z_{\mathrm{off}}=1\,$nm, resulting in an average force, over one oscillation, of 30\,nN when $V_{\mathrm{tip}}=2\,$V.
Following \cite{Hembacher2004}, the effect of removing this bias is that the average tip-sample distance 
would increase by only 17\,pm.
Thus, even accounting for the effect of removing the bias voltage, we needed to approach the tip over 300\,pm to the surface to observe the $\Delta f$ contrast in Fig.~\ref{stm2}(c).
It is therefore not possible that the frequency shift increase (e.g. in Fig.~\ref{si111} or Fig.~\ref{stm2}(a)) was recorded in region I where Pauli repulsion dominates.

We now consider the effect of the sample having a resistance given by $R_s$, with the tip biased at  $V_{\mathrm{tip}}$.
The electrostatic interaction between tip and sample causes a force that scales as the square of the voltage difference, $(V_{\mathrm{tip}} - V_{\mathrm{sample}})^2$.
We neglect the contact potential difference, $V_{\mathrm{cpd}}$, because it is simply an offset of the applied bias voltage.
While  local variations in $V_{\mathrm{cpd}}$ have been reported on the \sev surface with Kelvin probe measurements~\cite{Sadewasser2009}, 
they would not lead to an increase in $\Delta f$ over adatoms in both bias polarities.  
Letting $K$ represent the prefactor which is independent of bias~\cite{Hudlet1998}: 
\begin{equation}
F^{es} = -K\,(V_{\mathrm{tip}}^2 - 2 \, V_{\mathrm{tip}} \, I\,R_s + I^2\,R_s^2)
\end{equation}
Assuming $V_{\mathrm{sample}}=I\,R_s \ll V_{\mathrm{tip}}$, the second order term can be discarded.
Proceeding with the derivative to k$_{ts}$, where $I=I_0\,\mathrm{exp}(-\kappa z)$:
\begin{equation}
k_{ts}^{es} \equiv -\frac{\mathrm{d}F^{es}}{\mathrm{d}z} = \frac{\mathrm{d}K}{\mathrm{d}z}\, V_{\mathrm{tip}}^2 - (2 \,\frac{\mathrm{d}K}{\mathrm{d}z} - 2\,K\, \kappa)\, V_{\mathrm{tip}} \, R_s\, I
\end{equation}
$\Delta f$ is, in the small amplitude approximation, directly proportional to $k_{ts}$, and if $k_{ts}^{es}$ is dominant, $\Delta f$ would have a linear response with respect to $I$ and a slope that is linearly proportional to $V_{\mathrm{tip}}$.

%%%%%%%%%%%%%%%%%%%%%%%%%%%%%%%%%%%%%%%%%%%%%%%
\begin{figure}%[ht]
\begin{center}
\includegraphics[width=\columnwidth]{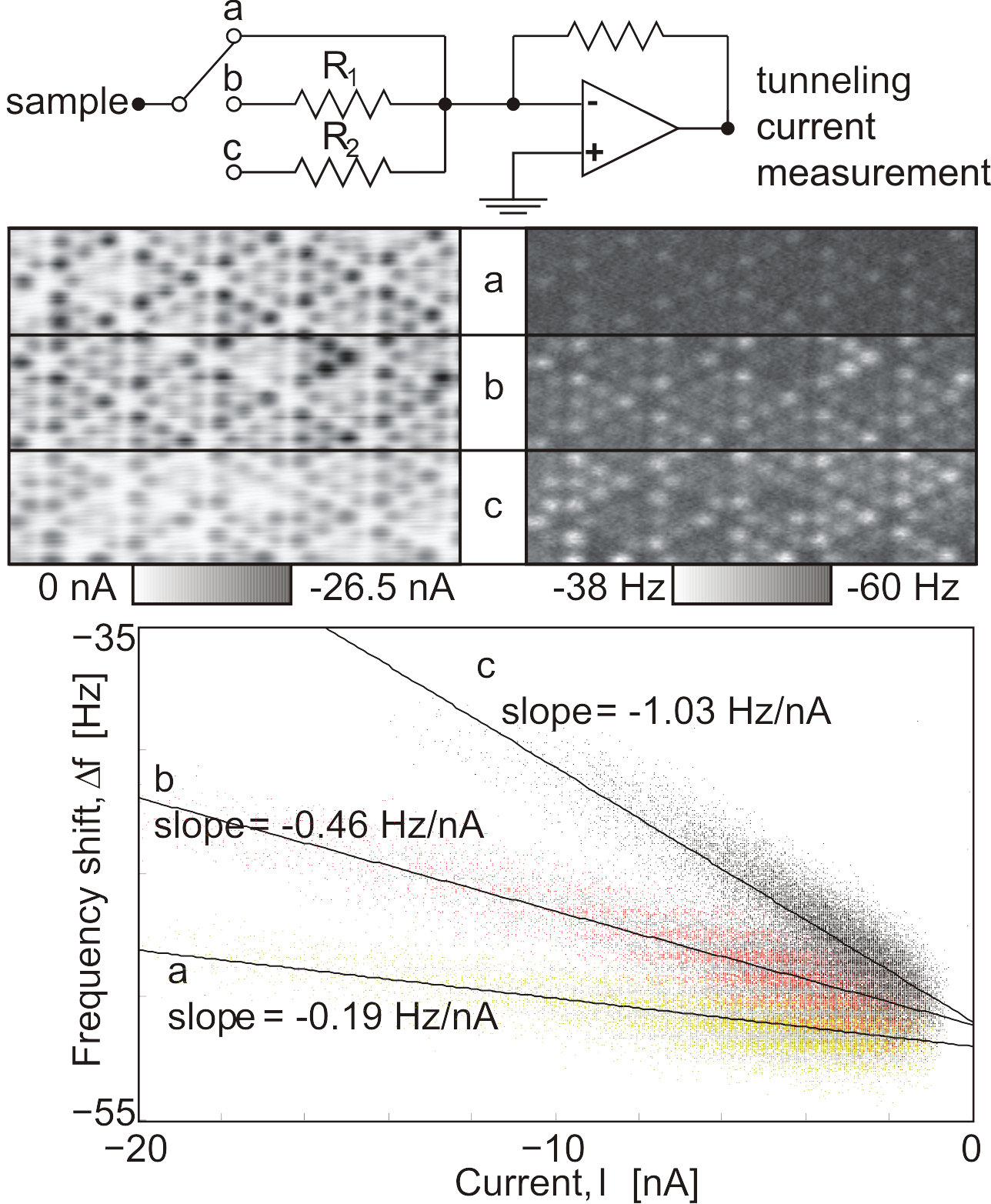}
\caption{
\textit{Top:} Schematic of the experiment.
\textit{Middle:} Simultaneous $I$ and $\Delta f$ data.  
\textit{Bottom:} $\Delta f$ versus $I$ data.
a) Data was acquired with no additional resistor between sample and virtual ground.  b) $R_1=10\,$M$\Omega$ was inserted between sample and ground.  c) $R_2=30\,$M$\Omega$ was inserted.  See text for details.  Data were acquired with $A=400$\,pm, $f_0=19\,390\,$Hz and images are 10\,nm$\times$7\,nm.
}
\label{oscdefl}
\end{center}
\end{figure}
%%%%%%%%%%%%%%%%%%%%%%%%%%%%%%%%%%%%%%%%%%%%%%%

To test this theory, we performed an experiment shown schematically in the top third of Fig.~\ref{oscdefl}.
A switch was installed between the high-doped sample and the virtual ground. 
The virtual ground is provided by the operational amplifier used to detect the tunneling current.
The switch is used to add known resistances $R_1=10$\,M$\Omega$ and $R_2=30$\,M$\Omega$.
$\Delta f$ should now follow the relation:
\begin{equation}
\Delta f =b + m\,V_{\mathrm{tip}}\,( R_s +  R) \,I
\label{dfvsRI}
\end{equation} 
where $R$ is either 0, $R_1$ or $R_2$, depending on the switch, $m$ is independent of $R_s$, V$_{\mathrm{tip}}$ and $I$, and $b$ is independent of $I$.
Considering the $I$-dependent term in Eq.~\ref{dfvsRI}, $m$ and $R_s$ are unknown.
The data collected at $R=0$ and $R=R_1$ can be used to determine $m$ and $R_s$, and then to predict a slope of -0.99 Hz/nA when $R=R_2$. 
The data and the slopes for the three switch settings are shown in Fig.~\ref{oscdefl}; the observed slope when $R=R_2$ is -1.03 Hz/nA.

This strong agreement with our simple model, incorporating a sample resistance $R_s$, prompts us to further explain the mechanism of this model.
Our fit parameters, for example, indicate that for the low-doped sample $R_s=164$\,M$\Omega$.
While this number seems high, the tunneling current is injected in a very small area, and that this in effect creates a large current density that must disperse through a relatively small area.  
Assuming that the current $I$ is injected in an area of radius 1\,\AA, and then disperses through the sample radially, we can use classical electrodynamics to determine the order of magnitude of the expected resistance.
The current density then scales as $1 / (2\,\pi\,r^2)$, where $r$ is the distance from the current injection,
and is directly proportional to the derivative of the electrochemical potential via the sample resistivity.
The voltage between the area in which the tunnel current is being injected and a position in the bulk sufficiently far is thus
\begin{equation}
V_{\mathrm{sample}} =  R_s \,I = \frac{ \rho }{2\,\pi\,(1\,\mathrm{\AA})}\, I
\end{equation}
Given this sample has a resistivity  $\rho=6$ to $9\,\Omega\,$cm, $R_s$ is in the range of 96 to 143\,M$\Omega$, which agrees well with the fitted $R_s$ value.
In this simple picture, the voltage drop would be highly local, but the charge density and electric field are both most intense near the tip apex, where one would expect the highest contribution to the electrostatic attraction.
This picture also has a much higher error when we consider the high-doped sample, and it is likely that an atomic-scale theory of electronic conductance is required to fully explain the observed $R_s$.  

To summarize, we observe a frequency shift increase over adatoms that scales linearly with current. 
We have characterized this response to current as a function of bias voltage.
This frequency shift can be explained by the effect that the tunnel current has on the surface potential, given non-negligable sample resistivity.

As far back as simultaneous FM-AFM and STM have been attempted, 
contrast inversion has been observed as function of tip state and of applied bias voltage~\cite{Molitor1999, Arai2000, Arai2000a}.
Guggisberg and coworkers suggest that short-range electrostatics might explain the contrast inversion, but do not propose a model for this~\cite{Guggisberg2001}.
Contrast inversion can be explained within our model quite easily: 
At low biases, the tip images the adatoms as predicted by theory, and shows a negative $\Delta f$ contrast over adatoms, while at higher biases, the $\Delta f$ contrast is due primarily to the decrease in capacitive attraction due to the sample resisitance and tunnel current, as reported in this Letter.

High spatial resolution demands short-range forces, which implies small tip-sample distances.
Given an interest in simultaneous AFM and STM, one must be aware that a moderate tip-sample bias can also produce a tunnel current so large that it would affect the surface potential and measurable $\Delta f$ contrast.
For surfaces such as Cu, with a $\rho \approx 2\times10^{-8}\,\Omega$\,m, this effect is ignorable~ \cite{Ternes2011}; even at V$_{\mathrm{tip}}=10$\,V and $I=100\,$nA, the expected $\Delta f$ would be $<1\,$mHz.
However this effect must be taken account of when performing combined STM/AFM of any surface with appreciable resistivity, including, of course, semiconductor surfaces.  

We thank J.~Repp, M.~Emmrich and M.~Schneiderbauer for discussions and J.~Mannhart for support.

\end{document}